\documentclass[pra,twocolumn,showpacs,preprintnumbers,amsmath,amssymb,nofootinbib,epsfig,bmamsfonts,yfonts]{revtex4}
\usepackage{graphicx}

\begin{document}

\newtheorem{lemma}{Lemma}
\newtheorem{theorem}{Theorem}
\newtheorem{corollary}{Corollary}
\newtheorem{definition}{Defnition}

\title{Few-body hierarchy in non-relativistic functional renormalization group equations and a decoupling theorem}


\author{Stefan Floerchinger}


\affiliation{Physics Department, Theory Unit, CERN, CH-1211 Gen\`eve 23, Switzerland\\
                E-mail: stefan.floerchinger@cern.ch}

\begin{abstract} 
For non-relativistic quantum field theory in the few-body limit with instantaneous interactions it is shown within the functional renormalization group formalism that propagators are not renormalized and that the renormalization group equations of one-particle irreducible vertex functions are governed by a hierarchical structure. This hierarchy allows to solve the equations in the $n$-body sector without knowledge or assumptions about the $m$-body sectors where $m>n$.
\end{abstract}

\maketitle

\section{Introduction}
\label{sec:1}

In classical mechanics as well as non-relativistic quantum mechanics it is an evident fact that one can solve the equations governing $n$ interacting massive particles without having to have any knowledge or making any assumptions about the solution to the corresponding equations for $n+1$ or more particles. This important and fundamental feature is not given for relativistic quantum field theory, however. Indeed, the non-perturbative renormalization group equations for the propagator (or two-point function) usually depends on the four-point function. The renormalization group equation of the four-point function, in turn, depends on the six-point function and so on. Strictly speaking, this infinite hierarchy of coupled equations makes it impossible to solve the equations governing a single particle, i.\ e.\ the equation for the propagator without considering the problem of two or more particles at the same time. (The situation is different in perturbation theory at fixed finite order.)

The decoupling feature is also broken by effects of non-zero density and temperature in traditional (non-relativistic) many-body theory. In the context of nuclear matter, it was shown by Br\"uckner, Bethe, Goldstone and others in the 1950's that the interaction with the Fermi sea of all particles leads to corrections of the self-energies and the interaction potentials between particles, see for example \cite{FetterWalecka}. 

In this paper we review the general arguments for the decoupling between $n$-particle problems in non-relativistic quantum field theory and discuss explicitly how it is realized in the formalism of the functional renormalization group. More specific we consider field theories which have only particles but no anti-particles or holes. The prime example for this is non-relativistic quantum field theory with instantaneous interactions in the few-body limit but there exist also other examples such as statistical reaction-diffusion systems \cite{Benitez:2012ax}. We will show that the renormalization group equations of such a field theory are governed by an interesting hierarchical structure which we call $n$-body hierarchy. This hierarchical structure allows in principle to solve the renormalization group equations for correlation functions governing $n$ interacting particles successively, i.\ e.\ first the equations for two particles, then for three particles and so on.

In the case of non-relativistic quantum field theory in the few-body limit one has an intuitive reason to expect such a structure. Indeed, this field theory is expected to be equivalent to non-relativistic quantum mechanics. The hierarchical structure on the field theory side corresponds on the side of quantum mechanics to the fact that one can solve the Schr\"odinger equation for the problem of $n$ interacting particles without having to solve the equations for $n+1$ particles. 

Historically, quantum many body theory and its field theoretic formulation have been developed by starting from the quantum mechanical formalism for many particles. In the formalism of second quantization the decoupling feature corresponds to the fact that the effective Hamiltonian in the few-body limit does not mix parts of the Fock space corresponding to different particle numbers. It can be proven using Wick's theorem.
In diagrammatic perturbation theory one can see that certain classes of diagrams, in particular all self energy corrections, vanish, see for example \cite{FetterWalecka}. We briefly review the corresponding arguments in section \ref{sec2}. The main goal of the present paper is to discuss the few-body hierarchy and its implications within the non-perturbative functional renormalization group formalism.

We consider here non-relativistic field theories with conserved particle number. In a field theoretic formulation this conservation law is connected with a global $U(1)$ invariance for each conserved particle species. In this paper we concentrate for simplicity on the case of a single particle species with bosonic quantum statistics and without spin. A generalization of the result to more complicated situations with different species, spin and fermonic statistics is straight forward.

We assume that the microscopic inverse propagator is of the form
\begin{equation}
i p_0 + f(\vec p) - \mu,
\label{eq:001}
\end{equation}
where $p_0$ is an euclidean (or imaginary) frequency and $\mu$ is a chemical potential. In the case equivalent to non-relativistic quantum mechanics one has $f(\vec p) = \frac{1}{2M}\vec p^2$, for a reaction diffusion system $f(\vec p) = D \vec p^2$ but actually the precise form of $f(\vec p)$ is not important for our purpose. We only require $f(\vec p)\geq 0$. The chemical potential $\mu$ is chosen such that the particle density vanishes. This implies in any case $\mu \leq 0$ and if there are bound states in the system one has to choose $|\mu|$ such that it is larger than the maximal binding energy per particle. This ensures that for fundamental particles as well as for composite particles or bound states the dispersion relation is always such that the on-shell energy $E=- i p_0$ is positive or zero. Moreover, $|\mu|$ must be chosen large enough that possible branch cuts for composite particle propagators are at positive energy.

Now that we have specified the microscopic propagator let us turn to the interactions. We assume that they are instantaneous which implies that the Fourier transformed microscopic interaction vertices are independent of frequency. In contrast we make no assumptions about the dependence of the interaction energy on the distance between particles. In Fourier space the dependence on the spatial momentum remains therefore unspecified. Also, we allow in addition to two-particle interactions interaction terms involving three or more particles. The only condition is that these must be instantaneous, as well.

\begin{figure}
\centering
\includegraphics[height=0.1\textwidth]{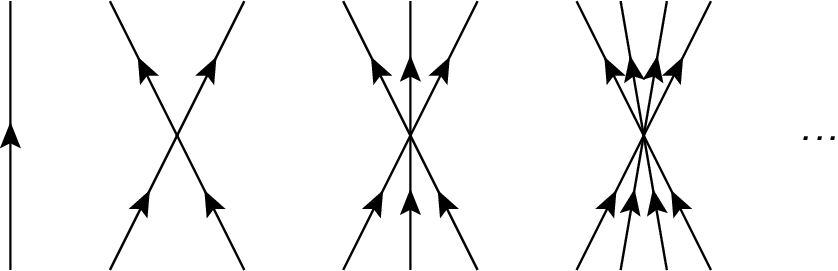}
\caption{Graphical representation of the propagator and microscopic interaction vertices.}
\label{fig:001}
\end{figure}

In a graphical notation we denote particles with the dispersion relation \eqref{eq:001} by a solid line with an arrow which denotes the direction of particle number flow. The microscopic two-particle interaction is denoted by a vertex where two lines cross, the three-particle interaction by a vertex where three lines cross etc., see Fig.\ \ref{fig:001}.

This paper is organized as follows. In section 2 we recall briefly some aspects and implications of the few-body hierarchy in perturbation theory. The material presented there is not new, it is included mainly to introduce the notation and to set the stage for the subsequent discussion of the non-perturbative renormalization group equations. Section 3 contains a classification of correlation functions that is useful later on. In section 4 we give a brief account of the functional renormalization group formalism and discuss some of its technical aspects that are important for our purpose, in particular related to the choice of an appropriate infrared regulator function. The implications of the few-body hierarchy for the functional renormalization group formalism are formulated as a decoupling theorem in section 5. The subsequent section 6 contains its proof and we draw some conclusions in section 7.

\section{Perturbative loop expansion}
\label{sec2}
From the conditions on the theory formulated in the previous section one can derive some interesting properties and relations. In this introductory section we concentrate on perturbation theory and its diagrammatic representation in terms of Feynman diagrams. The statements made here are not new and can be found in similar form at various places in the literature. They are presented here nevertheless in order to introduce the notation and as a warm-up for the subsequent discussion within the functional renormalization group formalism.

We start with a simple but powerful consequence of the form of the microscopic propagator. Since we will show below that the propagator is not renormalized at any order in perturbation theory or non-perturbatively, one can actually generalize the statements accordingly.

\begin{lemma}\label{lem:001}
The microscopic propagator as a function of the (imaginary) time difference $\Delta \tau$ between initial and final state vanishes when $\Delta \tau <0$. 
\end{lemma}
For the proof we employ the Fourier representation
\begin{equation}
G(\Delta \tau) = \int \frac{d p_0}{2 \pi} \frac{1}{i p_0 + f(\vec p)-\mu} e^{i p_0 \Delta \tau}.
\end{equation}
For $\Delta \tau>0$ one can close the $p_0$ integration contour in the upper half of the complex plane. The propagator has a pole there and the result is non-zero. However, for $\Delta \tau<0$ one has to close the integration contour in the lower half plane and the result vanishes. All particles must therefore propagate forwards with respect to the imaginary time direction $\tau$.

One can immediately derive some interesting consequences.
\begin{corollary}\label{cor:001}
All non-vanishing diagrams involving microscopic instantaneous interaction vertices can be drawn such that there is a preferred direction ("time") in which incoming and outgoing lines point and no particle flows backwards against this direction.
\end{corollary}
This is directly evident from lemma \ref{lem:001}.

\begin{corollary}\label{cor:002}
To all orders in perturbation theory the (inverse) propagator in \eqref{eq:001} is not renormalized.
\end{corollary}

Indeed particle number conservation implies that all possible diagrams renormalizing the two-point function contain a part where particle number flow points against the time direction and vanishes therefore due to the above statements.

\begin{corollary}\label{cor:002}
All diagrams in perturbation theory that contain a closed loop of particle flow vanish.
\end{corollary}

\begin{figure}
\centering
\includegraphics[height=0.1\textwidth]{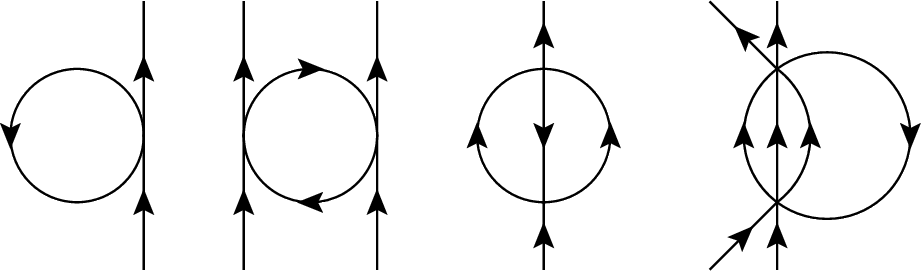}
\caption{Feynman diagrams that contain a closed loop of particle number flow and vanish since they necessarily contain lines pointing backwards in time (downwards in the graphical representation). The interaction vertices are microscopic, instantaneous interactions.}
\label{fig:002}
\end{figure}

Indeed, a closed loop where all particle number flow arrows point along the loop direction necessarily contains a part where particle flow lines point backwards in time and vanishes therefore. Some examples for this are shown in Fig.\ \ref{fig:002}. A statement similar to this one was proven in ref.\ \cite{Diehletal}.

\section{Classification and decomposition of correlation functions}
After the brief introductory discussion of perturbation theory let us streamline the discussion a bit. The main objective of this section is to classify correlation functions. We define terms such as ``$n$-body" and ``$m$-closeable" that will be useful below. Although we use perturbation theory in some of the arguments, the statements and in particular the classification of correlation functions are also valid beyond perturbation theory. For a formulation that does not rely on Feynman diagrams one works in the functional integral formulation and introduces appropriate source terms of linear, quadratic and higher order in the fundamental fields.

\begin{figure}
\centering
\includegraphics[height=0.1\textwidth]{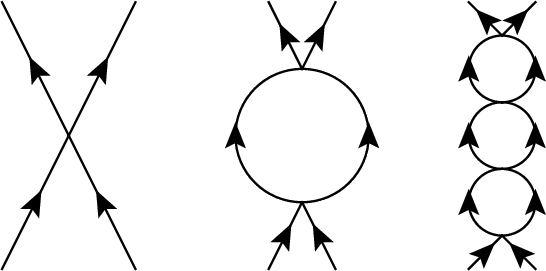}
\caption{Two-body diagrams.}
\label{fig:003}
\end{figure}

\begin{figure}
\centering
\includegraphics[height=0.1\textwidth]{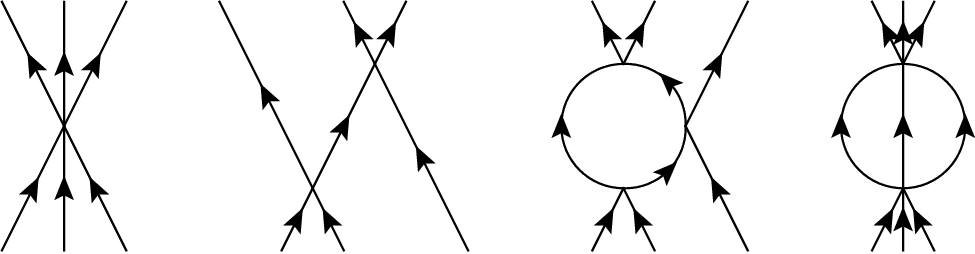}
\caption{Three-body diagrams.}
\label{fig:004}
\end{figure}

Let us start with the following
\begin{definition}
We call a (connected) diagram \emph{$n$-body} if it contains $n$ incoming and $n$ outgoing lines. 
\end{definition}
Due to the non-renormalization property for the propagator there are no non-trivial one-body diagrams in perturbation theory. In Fig.\ \ref{fig:003} we show some examples of two-body diagrams and some three-body diagrams are shown in Fig.\ \ref{fig:004}.

Note that all possible connected two-particle diagrams are one-particle irreducible. Apart from the tree-level diagram they are all two-particle reducible, however. This property signals already that a one-particle irreducible scheme might be particularly useful  in the two-body sector and indeed one can close and solve the corresponding renormalization group equations there, as will be discussed in more detail below.

The class of three-body diagrams contains also one-particle reducible tree-level diagrams. The loop contributions can be one-particle and two-particle irreducible but are all three-particle reducible. Obviously these properties can be generalized to $n$-body graphs:
\begin{lemma}\label{lem:002}
An $n$-body correlation function involving more than a single fundamental interaction vertex is at most $(n-1)$ particle \emph{irreducible} but $n$-particle \emph{reducible}.
\end{lemma}

For the proof it suffices to note that due to particle number conservation and corollary \ref{cor:001} at each time step there are always $n$ forward propagating lines connecting the incoming and outgoing lines. One can always make a horizontal cut between two vertices showing $n$-particle reducibility. On the other side one can always construct $(n-1)$-particle irreducible graphs, for example the generalizations of the last diagram in Fig.\ \ref{fig:004}.

Let us now define a property of $n$-body correlation functions that will be particularly useful for studying the renormalization group evolution.

\begin{definition}
We call a $n$-body diagram \emph{$m$-closeable} if $m$ outgoing lines can be connected to ingoing lines such that a non-vanishing $(n-m)$-body diagram is obtained.
\end{definition}

As an example consider the second diagram in Fig.\ \ref{fig:004}. It is one-closeable since one can connect the outgoing line on the left with the incoming line on the right. The result is the one-loop two-body diagram in Fig.\ \ref{fig:003}.

\begin{figure}
\centering
\includegraphics[height=0.1\textwidth]{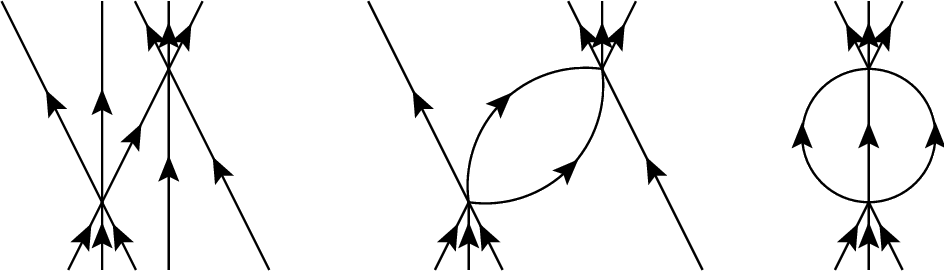}
\caption{The two-closeable $5$-body diagram on the left can be closed to yield the one-closeable $4$-body diagram in the middle. Closing it once more yields the two-loop three-body diagram on the right.}
\label{fig:005}
\end{figure}

As a second example consider the $5$-body diagram in Fig.\ \ref{fig:005}. It can be closed twice to yield a non-vanishing three-body diagram and is therefore two-closeable. 

We note that one can determine from the analytic structure of an $n$-point function with respect to the frequency of the incoming and outgoing lines whether it is closeable or not. Indeed, as a function of the potential loop frequency it must contain a singularity (a pole or a branch-cut) in the lower half of the complex plane. Otherwise one could close the integral contour there and the resulting loop expression would vanish.

In the (imaginary) time representation a $n$-body diagram is $m$-closeable when it is non-vanishing for a configuration where the time arguments for $m$ outgoing lines $t_\text{out}^1,\dots,t_\text{out}^m$ are earlier than the time arguments of $m$ incoming lines $t_\text{in}^1,\dots,t_\text{in}^m$. More precisely one needs a non-vanishing value for a configuration with 
\begin{equation}\label{eq:mcloseabletime}
t_\text{out}^1<t_\text{in}^{P(1)},\;\;\dots,\;\; t_\text{out}^m < t_\text{in}^{P(m)},
\end{equation}
where $P(1),\;\dots,\;P(m)$ is a permutation of $1,\;\dots,\;m$.

This shows that the attribute $m$-closeable is not only well defined for a perturbative contribution to an $n$-body correlation function but can be used more generally to characterize these objects also beyond perturbation theory.

We remark that the non-renormalization property of the propagator implies that two-body correlation functions are never closeable. Similarly, three-body correlation functions can be at most one-closeable. Also, since two-body correlation functions with loops are one-particle irreducible but two-particle \emph{reducible}, opening a line leads to a one-particle reducible three-body correlation function. Closeable three-body correlation functions are therefore one-particle reducible. The generalization of these statements leads to the following bound:
\begin{lemma}
The one-particle irreducible $n$-body vertex function can have $m$-closeable parts where
\begin{equation}
m\leq n-3.
\end{equation}
\end{lemma}
Indeed, closing the $m$ lines yields a $(n-m)$-body correlation function. The latter must be at least 3-body since opening a line of a two-body correlation function yields a one-particle reducible diagram.

To conclude this section let us remark that the classification of an $n$-body diagram being closeable is related to a classification proposed originally by Taylor \cite{Taylor}, adapted and simplified for time-ordered perturbation theory by Thomas and Rinat \cite{ThomasRinat} as well as Afnan and Blankleider \cite{AfnanBlankleider} and reviewed by Phillips and Afnan \cite{PhillipsAfnan}. In these works, a non-standard definition of $n$-particle irreducibility is used where one demands that possible cuts are horizontal lines that separate all incoming from all outgoing particles. On the other side, the cuts may include incoming and outgoing lines in addition to at least one internal line. 
For the class of theories considered here (with conserved number of particles and no anti-particles), connected $n$-body diagrams (or contributions to $n\to n$ Greens functions) allow then only for so-called $n$-cuts and are therefore at least $(n-1)$-particle irreducible. According to Taylors classification, they can be placed in one of the three classes: $C_1$ (no $n$-cut is possible, i.e. $n$-particle irreducible diagrams), $C_2$ (at least one $n$-cut is possible but it necessarily involves only internal lines) and $C_4$ (at least one $n$-cut can be made that involves also incoming and outgoing lines). A diagram is closeable precisely when it is in the class $C_4$ since one can then connect an outgoing to an incoming line.

\section{Non-perturbative renormalization group formalism}
In this section we discuss the non-perturbative renormalization group evolution of $n$-body correlation functions in a one-particle irreducible scheme. We introduce briefly the functional renormalization group formalism and the particular adaptions that must be made to apply it in our situation (such as choosing an appropriate infrared regulator function). For more detailed introductions and for reviews we refer to the literature \cite{RGReviews}. 

The formalism is based on the functional integral formulation of quantum field theory. In our situation the (euclidean) microscopic action is of the form
\begin{equation}
\begin{split}
S[\phi]=\int d\tau \int d^{d-1}x & {\Big \{} \phi^*\left(\partial_\tau + f(-i \vec \nabla)-\mu\right) \phi \\
& + \text{interaction terms}  {\Big \}}
\end{split}
\label{eq:micrscopicaction}
\end{equation}
where the interaction terms are of quadratic and higher order in the fields $\phi^*$ and  $\phi$. Since we assume interactions to be instantaneous they involve no derivatives with respect to the (imaginary) time $\tau$.

In the non-perturbative renormalization group formalism one adds to this a term quadratic in the fields $\phi^*$ and $\phi$. In momentum space it reads
\begin{equation}
\Delta S_k = \int \frac{dp_0}{2\pi} \int \frac{d^{d-1}p}{(2\pi)^{d-1}} \phi^* R_k(p_0,\vec p) \phi.
\end{equation}
Here the index $k$ denotes an momentum scale and $R_k$ is chosen such that it acts as an infrared regulator at that scale. In general it is advisable to choose $R_k(p_0,\vec p)$ such that it does not break any important symmetries present for the microscopic action \eqref{eq:micrscopicaction}, although this is not strictly necessary (an alternative is to consider Ward identities modified by the explicit symmetry breaking due to $R_k$).

Except from this one usually requires that
\begin{equation}\label{eq:uvlimitcutoff}
R_k(p_0,\vec p) \to 0 \quad \text{for} \quad k\to 0
\end{equation}
and
\begin{equation}\label{eq:irlimitcutoff}
R_k(p_0,\vec p) \to \infty \quad \text{for} \quad k\to \infty.
\end{equation}
A simple choice fulfilling these properties is $R_k(p_0,\vec p) = k^2$. In particular for approximate calculations it is furthermore often useful to work with a cutoff function that decays for large momenta and frequencies.

In our specific context there is one further crucial requirement. As discussed earlier, the singularities of the propagator
\begin{equation}
G=\frac{1}{i p_0 + f(\vec p)-\mu}
\end{equation}
with respect to $p_0$ are all in the upper half of the complex plane. It will be important for what follows that this is also the case for the regularized propagator
\begin{equation}
G_k=\frac{1}{ip_0 + f(\vec p)-\mu + R_k(p_0,\vec p)}.
\end{equation}
Obviously the cutoff $R_k=k^2$ fulfills this requirement as well as any other choice with $Re(R_k)\geq 0$ that depends only on the spatial momentum $\vec p$.

At this point we mention only one further possible choice \cite{SFAnalyticContinuation} that decays for larger frequencies and momenta and is furthermore invariant under the Galilean symmetry present for $f(\vec p)=\frac{1}{2M}\vec p^2$ ($c>0$ is an arbitrary real parameter, for concreteness take $c\approx 1$),
\begin{equation}
R_k(p_0,\vec p) = \frac{c k^4}{c k^2 + i p_0 + f(\vec p) - \mu}.
\end{equation}
Indeed, with this choice the regularized propagator has poles for
\begin{equation}
ip_0+f(\vec p)-\mu = \left(-\frac{c}{2} \pm i \sqrt{c-\frac{c^2}{4}}\right) k^2.
\end{equation}
For most of the discussion that follows here it will not be necessary to choose a specific form of the regulator function as long as all of the above requirements are fulfilled. 

From this point one proceeds by defining a modified version of the Schwinger functional, the generating functional for connected correlation functions,
\begin{equation}
e^{W_k[J]} = \int d\phi \; e^{-S[\phi]-\Delta S_k[\phi] + \int_{\tau,\vec x}\{J^*\phi + \phi^* J\}}.
\end{equation}
The Legendre transform of this yields the generating functional of one-particle irreducible correlation functions in the presence of the infrared regulator term $\Delta S_k$,
\begin{equation}
\tilde \Gamma_k[\phi] = \int_{\tau,\vec x} \{J^*\phi + \phi^* J\} - W_k[J]
\end{equation}
where the right hand side is evaluated for
\begin{equation}
\phi = \frac{\delta}{\delta J^*} W_k[J], \quad \phi^* = \frac{\delta}{\delta J} W_k[J].
\end{equation}
Subtracting from this the infrared regulator function yields the flowing action,
\begin{equation}
\Gamma_k[\phi] = \tilde \Gamma_k[\phi] - \Delta S_k[\phi].
\end{equation}
Due to the property \eqref{eq:uvlimitcutoff} of the infrared regulator function one recovers the standard generating functional for one-particle irreducible correlation functions for $k\to 0$,
\begin{equation}\label{eq:uvlimitGamma}
\lim_{k\to 0} \Gamma_k[\phi] = \Gamma[\phi].
\end{equation}
On the other side, the property \eqref{eq:irlimitcutoff} implies that all quantum fluctuations are suppressed for large infrared cutoff scales. This implies
\begin{equation}\label{eq:irlimitGamma}
\lim_{k\to \infty} \Gamma_k[\phi] = S[\phi].
\end{equation}

Let us now come to the central equation of the functional renormalization group formalism, the exact flow equation \cite{CWFlowEquation} for $\Gamma_k[\phi]$,
\begin{equation}\label{eq:FlowEquation}
\partial_k \Gamma_k[\phi] = \frac{1}{2} \text{Tr} (\Gamma_k^{(2)}[\phi]+R_k)^{-1} \partial_k R_k.
\end{equation}
Here $\text{Tr}$ is the trace operator. Since Eq.\ \eqref{eq:FlowEquation} can be used to follow the changes in $\Gamma_k[\phi]$ between the two limiting cases in Eqs.\ \eqref{eq:uvlimitGamma} and \eqref{eq:irlimitGamma}, the exact flow equation in \eqref{eq:FlowEquation} allows to take the effect of quantum fluctuations on the generating functional $\Gamma_k[\phi]$ into account. Note that although \eqref{eq:FlowEquation} is of a relatively simple one-loop structure it is nevertheless non-trivial since both sides depend on the fields $\phi^*$, $\phi$ in a functional way.

For our purpose it will also be useful to work with the following version of the exact flow equation
\begin{equation}\label{eq:FlowEquation2}
\partial_k \Gamma_k[\phi] = \tilde \partial_k \frac{1}{2} \text{STr} \ln (\Gamma_k^{(2)}[\phi]+R_k)
\end{equation}
where $\tilde \partial_k$ is a formal derivative that hits only the cutoff $R_k$. In this form the one-loop form is even more evident. In particular the functional derivatives of \eqref{eq:FlowEquation2} lead to flow equations for correlation functions that resemble the perturbative diagrams. The vertices are replaced by complete frequency-, momentum- and $k$-dependent vertices and for the propagator one has to take a variant regularized by the infrared cutoff $R_k$.

\section{The decoupling theorem and its implications}
In this section we present and discuss a useful decoupling theorem which essentially states that the renormalization group equations governing the $n$-body sectors can in principle be solved succeedingly, starting from $n=2$, then going to $n=3$ and so on. The proof of the theorem is given in the following section.

As shortly discussed in section \ref{sec:1}, the presence of a hierarchical structure and a decoupling of $n$-body sectors for the class of theories considered here does not come as a surprise. In the canonical or operator formalism representation of quantum field theory, it can be understood as a property of the few-body Hamiltonian which does not mix $n$-particle subspaces of the complete Hilbert space. In principle one could learn about the implications of the hierarchy for the functional renormalization group picture by tracing its fate through the derivations of the functional integral and the renormalization group. Here we rather follow a more direct approach which also makes it easier to transfer the results to other related situations.

Let us start from a field expansion of the flowing action $\Gamma_k[\phi]$. Due to the $U(1)$ symmetry only terms of equal order in $\phi^*$ and $\phi$ can appear. We write
\begin{equation}
\Gamma_k[\phi] = \Gamma_k^{\{0\}} + \Gamma_k^{\{1\}}[\phi] + \Gamma_k^{\{2\}}[\phi] + \Gamma_k^{\{3\}}[\phi]+\dots
\end{equation}
where $\Gamma_k^{\{0\}}$ is independent of the field $\phi$ (and can be drooped since it is not relevant in the following), $\Gamma_k^{\{1\}}[\phi]$ is of the order $\phi^*\phi$, $\Gamma_k^{\{2\}}[\phi]$ of order $\phi^{*2}\phi^2$ and so on. 

We furthermore decompose the terms $\Gamma_k^{\{n\}}$ for $n>3$ into a sum of the form
\begin{equation}\label{eq:decompositionGammanm}
\Gamma_k^{\{n\}}[\phi] = \sum_{m=0}^{n-3} \Gamma_k^{\{n,m\}} [\phi].
\end{equation}
The term $\Gamma_k^{(n,m)}[\phi]$ is defined such that its functional derivative yields the $m$-closeable part of the one-particle irreducible $n$-body vertex function.

As discussed above one can base the definition of $m$-closeable either on the analytic structure of a correlation function with respect to its frequency arguments or on the imaginary time representation according to the discussion around Eq.\ \eqref{eq:mcloseabletime}. 

At this point we remark also that for large cutoff scales $k\to\infty$ the $\Gamma_k^{\{n\}}[\phi]$ for $n\geq 2$ approach the fundamental interaction terms present in the microscopic action $S[\phi]$. As fundamental instantaneous interactions they are non-closeable so that one can infer
\begin{equation}
\lim_{k\to\infty} \Gamma_k^{\{n,m\}}[\phi] = 0\quad \text{for} \quad m\geq1.
\end{equation}
Together with \eqref{eq:irlimitGamma} this fixes the decomposition \eqref{eq:decompositionGammanm} for $k\to\infty$.

The fact that the two-body part $\Gamma_k^{\{2\}}[\phi]$ does not have a closeable part has an interesting consequence.
\begin{lemma}\label{lem:nonrenormalizationpropagator}
The propagator or one-body part of the effective action $\Gamma_k^{\{1\}}[\phi]$ is not renormalized.
\end{lemma}
We remark that this statement has first been show within a truncation of the functional renormalization group equations in \cite{Diehletal}. The proof is straight-forward. Due to the one-loop property of the flow equation and particle number conservation, the only possible renormalization of the propagator comes from the tadpole similar to the first diagram in Fig.\ \ref{fig:001}. Lemma \ref{lem:001} ensures that the microscopic propagator can only point forwards in time. We assume that the regulator $R_k$ is chosen such that this holds also for the regularized microscopic propagator as well as for its regulator scale derivative. However, since the two-body interaction vertex is non-closeable this implies that the right hand side of the flow equation vanishes initially (at the microscopic scale) and therefore at all $k$.

It is an important consequence of lemma \ref{lem:nonrenormalizationpropagator} that one can rely on the statement of lemma \ref{lem:001} not only at the microscopic scale but on all scales $k$. More specific, both the regularized propagator $G_k(\Delta \tau)$ and its regulator scale derivative $\tilde \partial_k G_k(\Delta\tau)$ vanish when $\Delta\tau<0$.

We now come to a central statement of this paper.
\begin{theorem}[Decoupling theorem]\label{the:001} 
The right hand side of the flow equation for the $m$-closeable part of the one-particle irreducible $(n+m)$-body vertex function can depend only on (regularized) propagator, the $q$-body vertex functions with $2\leq q\leq n$ and the part of the $(n+j)$-body function that is at least $j$-closeable where $0 < j \leq m+1$.
\end{theorem}

Before proving this theorem let us shortly discuss its implications and consequences. To this end we make the following
\begin{definition}
The \emph{one-body sector} consists of the propagator and the \emph{two-body sector} of the one-particle irreducible two-body vertex. The \emph{$n$-body sector} for $n>2$ consists of the non-closeable part of the one-particle irreducible $n$-body vertex functions and the $m$-closeable part of the $(n+m)$-body vertex functions.
\end{definition}

From theorem \ref{the:001} one can now immediately infer the following
\begin{corollary}
The flow equations for the $n$-body sector depend only on the vertex functions in the $j$-body sectors where $j\leq n$ but not on the vertex functions in the $j$-body sectors where $j>n$.
\end{corollary}

In other words, the flow equation for the $n$-body sector decouple from the higher body sectors. This leads to the important consequence that one can first solve the flow equations for the two-body sector, use this solution to solve the equations for the three-body sector and so on subsequently for the higher $n$-body sectors.

Note that the one-body and two-body sectors are finite in the sense that they contain a limited number of correlation functions. For this reason one can in practice use the implications of the few-body hierarchy  to solve the equations in these sectors explicitly. In contrast, the three-body and higher sectors consist of an infinite chain of vertex functions. Without further truncations one can in general not expect explicit analytic solutions there. For some purposes it may be legitimate to discard $m$-closeable contributions to the $(n+m)$-body correlation functions for large $m$. Not only are they typically of higher order in the coupling constants but also of higher (inverse mass) scaling dimension (compared to the ones with smaller $m$) and therefore typically less relevant at infrared fixed points of the renormalization group evolution. The corresponding $n$-body sector becomes then finite and it might be possible to solve the corresponding RG equations explicitly.

We note at this point that despite the above restrictions, exact results have also been obtained from functional RG equations in the three-body sector without further truncations \cite{FRGFewBody1,FRGFewBody2}. To that end a composite dimer field was introduced using a Hubbard-Stratonovich transformation. The resulting renormalization group formalism differs somewhat from the one discussed here since the effective action generates then diagrams that are one-particle irreducible with respect to the original fields but also with respect to the composite dimer field. It is an interesting open question whether this strategy is actually more general and could also lead to exact solutions in the four- and five-body sectors.

Let us finally note that once one has calculated the one-particle irreducible $n$-body vertex functions one can directly extract physical observables such as scattering matrix elements or binding energies of bound states. For example, a scattering amplitude for three-to-three particle scattering is obtained by summing the one-particle irreducible three-body correlation function from the functional derivative of $\Gamma[\phi]$ with tree-level expressions similar to the second diagram in Fig.\ \ref{fig:004}. The two-body vertex appearing in these expressions is the complete one-particle irreducible one obtained from $\Gamma[\phi]$ as well.

\section{Proof of the decoupling theorem}
Let us now formulate the proof of theorem 1. We start with an auxiliary formula. Consider a one-loop expression as it appears on the right hand side of the flow equation with $n$ incoming and $n$ outgoing external lines (a contribution to the flow of the $n$-body correlation function). We assume that the expression involves $V_j$ vertices with $j$ incoming and outgoing lines (i.\ e.\ $j$-body vertices). By counting lines one finds the constraint
\begin{equation}\label{eq:aux1}
n= \sum_{j=2}^\infty V_j (j-1).
\end{equation}
Since $V_j\geq 0$ is an integer one has $V_j=0$ for $j>n+1$. Moreover, $V_{n+1}\neq 0$ implies $V_{n+1}=1$ and $V_j=0$ for $j<n+1$. In this case the loop consists of a single vertex that starts and ends on the $(n+1)$-body vertex. The latter must therefore be one-closeable to yield a non-zero contribution. We have now already proven theorem 1 for the simplest case $m=0$.

Let us now consider $m>0$. From the generalization of \eqref{eq:aux1} it follows now that the highest order correlation function that can be involved is $(n+m+1)$-body. 

Assume now that the one loop expression contains at least one $(n+j)$-body vertex function where $1\leq j \leq m+1$. (Otherwise there is no non-trivial statement to prove.) Let us call this vertex A and let us group the remaining vertices in the loop into a structure B which constitutes an $(m-j+1)$-body correlation function, although in general one-particle reducible. We now distinguish different cases how the vertex A and the correlation function B are connected to form a one-loop expression.

\begin{figure}
\centering
\includegraphics[width=0.2\textwidth]{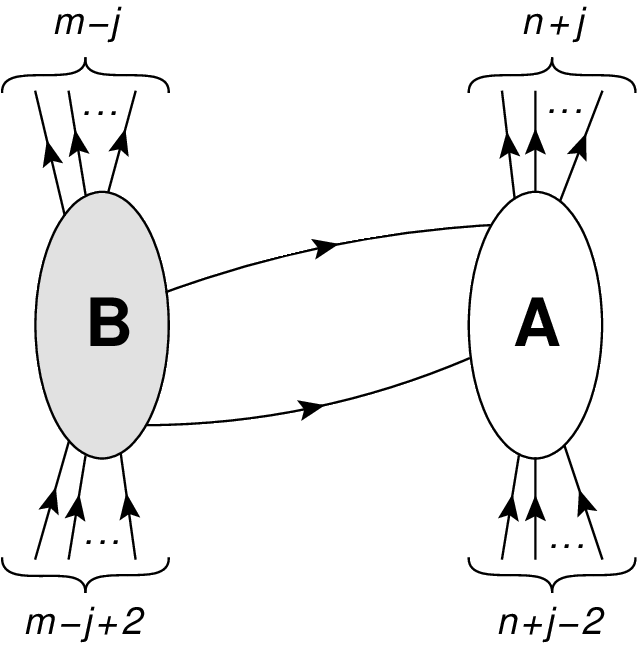}
\caption{Case (i).}
\label{fig:006}
\end{figure}

(i) The $(n+j)$-body vertex A is connected to the correlation function B by two lines that are incoming on A. For a graphical representation of this situation see Fig.\ \ref{fig:006}.

(ii) The $(n+j)$-body vertex A is connected to the correlation function B by two lines that are outgoing on A.

\begin{figure}
\centering
\includegraphics[width=0.2\textwidth]{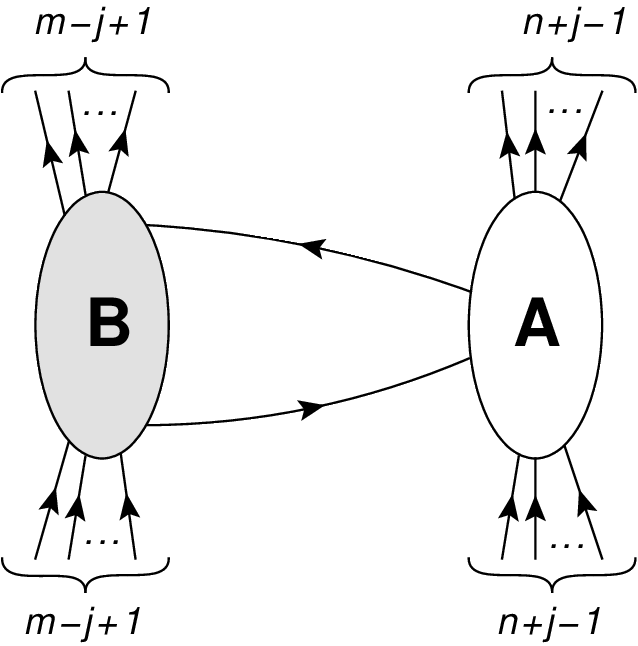}
\caption{Case (iii).}
\label{fig:007}
\end{figure}

(iii) The $(n+j)$-body vertex A is connected to the correlation function B by a line that is outgoing and one that is incoming on A. The time argument of the outgoing line is later than the one of the incoming line. This situation is shown graphically in Fig.\ \ref{fig:007}.

\begin{figure}
\centering
\includegraphics[width=0.2\textwidth]{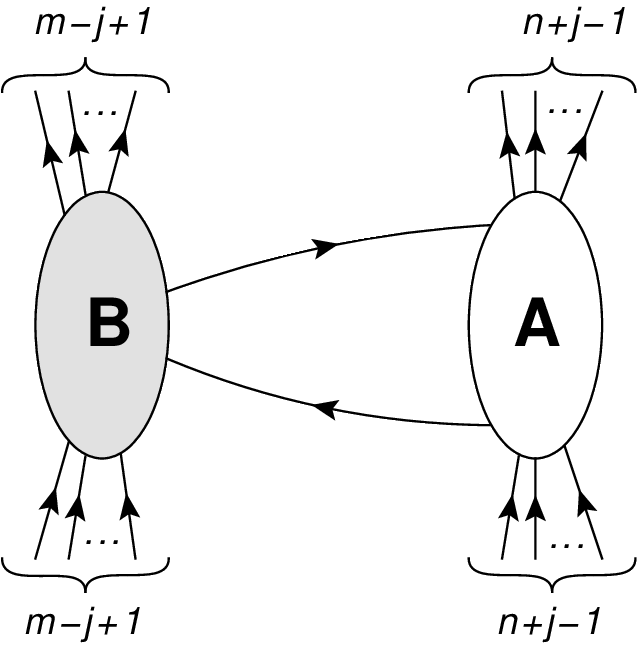}
\caption{Case (iv).}
\label{fig:008}
\end{figure}

(iv) As in (iii) but now the time argument of the outgoing line is before that of the incoming line. For a graphical representation of this situation see Fig.\ \ref{fig:008}.

We now consider what happens when one closes lines. In principle there are three possible ways. First, one can close a line by connecting an outgoing and an incoming line of the correlation function B without touching the vertex A. Second one can close lines such that one makes additional connections between the two objects and the third way is to close a line within the vertex A.

To prove that the vertex A must be $j$-closeable let us introduce the integer number $N_A$ counting how often we effectively close a line on A. We show that after closing $m$ lines in total, each of them either in the first, second or third way, one necessarily arrives at $N_A\geq j$. This in turn implies that the $(n+j)$-body vertex A must be at least $j$-closeable and therefore proves theorem 1.
The proof strongly relies on lemma \ref{lem:001} and lemma \ref{lem:nonrenormalizationpropagator} and we consider only diagrams that can be drawn such that all lines point forwards in time. 

Let us also introduce the "auxiliary counter" $N_B$ which count the number of available pairs of incoming and outgoing lines on the vertex B.

Let us start with case (i) (Fig.\ \ref{fig:006}). The auxiliary counter starts with $N_B=m-j$. Let us now attempt to close a line. If the line is closed in the first way one expends a pair of incoming and outgoing lines on B and and the auxiliary counter reduces by one, $N_B \to N_B-1$. Let us now attempt to close a line in the second way, i.e. by drawing a line between A and B. If one connects a line that is incoming on A with one that is outgoing on B there is always another line that incoming on B before that. Now there are two possibilities. Either this line remains an open incoming line even after all $m$ lines have been closed or it gets closed by connecting it with a line that is outgoing on A. In the former case we have simply closed one line and expended one pair of incoming and outgoing lines at B which lowers the auxiliary counter by one, $N_B\to N_B-1$. In the latter case we have closed two lines, and consumed one pair, i.\ e.\ we have to set $N_B\to N_B-1$. However, in effect the two lines together constitute a line that starts and ends on A and therefore one has to raise the corresponding counter $N_A\to N_A+1$. We note at this point that instead of starting with an outgoing line on B we could as well have started with an incoming line and connected it to an outgoing line of A. 

The steps described above can only be reiterated until the counter $N_B$ has reached $0$, i.\ e.\ at most $m-j$ times. The remaining lines must then be closed in the third way. Each step raises the counter $N_A\to N_A+1$. In total one finds at the end $N_A \geq j$ as required.

Case (ii) is obviously completely analogous to case (i).

Case (iii) is a little more intricate since we now have to start with $N_B=m-j+1$. However, one observes that there must always be an outgoing line with time argument after the loop line that is incoming on B. This remains to be true even after a number of lines have been closed in the first way on B. If this line remains to be an external line after closing all $m$ lines it reduces the counter $N_B$ effectively to $N_B=n-j$. Otherwise, if it gets connected with a line outgoing on A this constitutes together with the loop line a line that starts and ends on A and therefore to $N_A\to N_A+1$. From this point one can follow the argument as in case (i) leading in effect again to $N_A\geq j$.

In case (iv) one starts with $N_B=m-j+1$. Now, however, the two loop lines constitute already a line that starts and ends on A so that one has to initialize with $N_A=1$. The rest of the argument is then as in case (i).

This closes the proof of theorem 1.

\section{Conclusions}
We have discussed how the hierarchical structure of $n$-body sectors that is immanent in non-relativistic quantum mechanics is realized in the field theoretic functional renormalization group formalism. This is of interest due to several reasons:

(i) There is little doubt that non-relativistic quantum field theory in the few-body limit is equivalent to quantum mechanics. The discussion presented here helps to shed light on the precise details of this correspondence and thereby facilitates a comparison between results of (approximate) calculations in both formalisms.

(ii) Some of the arguments presented here for the case of non-relativistic few-body physics can be generalized to other quantum and statistical field theories with similar properties.

(iii) There are few cases known where one can obtain exact solutions of non-perturbative functional renormalization group equations. The mechanism that allows for this here is therefore of more general interest and its understanding might be of use in other situations as well.

(iv) In recent years the non-perturbative renormalization group formalism has been proven to be a useful tool for analytic and numerical investigations of universal few-body physics \cite{FRGFewBody1,FRGFewBody2,FRGFewBody3}. The theoretical insights obtained here will help to construct more effective and precise approximation schemes in the future.

\begin{acknowledgements}
I thank N.\ Wschbor for useful discussions and for making me aware of a gap in the previous understanding of the few-body hierarchy.
\end{acknowledgements}

\end{document}